\documentclass[preprint,12pt, a4paper]{elsarticle}

\usepackage{amssymb}
\usepackage{hyperref}
\usepackage{xcolor}
\usepackage{listings}

\newread\tmp

\journal{SoftwareX}

\begin{document}

\renewcommand{\labelenumii}{\arabic{enumi}.\arabic{enumii}}

\begin{frontmatter}



\title{rapidsim: A Python interface and cross-platform distribution framework for the RAPID protoplanetary disc simulation code}

\author[label1,label2]{D. Tarczay-Nehéz}
\address[label1]{Konkoly Observatory, HUN-REN CSFK, MTA Centre of Excellence, Budapest, Konkoly Thege Miklós út 15-17, Hungary, tarczaynehez.dora@csfk.org}
\address[label2]{MTA–HUN-REN CSFK Lendület ”Momentum” Stellar Pulsation Research Group}

\begin{abstract}
The \texttt{rapidsim} is a Python package for the \texttt{RAPID} (Representative Approach for Particle-Integrated Disks) numerical simulation code, which is a modular one-dimen\-sional hydrodynamic simulation code for modelling the evolution of protoplanetary discs and embedded dust particles written in C language. 
The package was developed to provide a simple and platform-independent tool for launching \texttt{RAPID} simulation without having to manually compile the native code and its external dependencies. 
\texttt{rapidsim} is distributed through the Python Package Index (PyPI) using pre-built binary wheels for Linux and macOS on both Intel and Apple Silicon architectures and for multiple Python versions. 
To ensure that the most recent package version is always available, an automated continuous-integration and release system synchronizes package versioning with \texttt{RAPID} releases, automatically building, validating, versioning, and publishing updated package releases.
\end{abstract}


\end{frontmatter}


\begin{table}[!h]
\begin{tabular}{|l|p{6.5cm}|p{6.5cm}|}
\hline
\textbf{Nr.} & \textbf{Code metadata description} & \textbf{Please fill in this column} \\
\hline
C1 & Current code version & v1.0.15 \\
\hline
C2 & Permanent GitHub link to code/repository used for this code version & \url{https://github.com/tnehezd/rapidsim} \\
\hline
C3 & Permanent link to Reproducible Capsule & \url{https://doi.org/10.5281/zenodo.23185394}\\
\hline
C4 & Legal Code License & MIT License \\
\hline
C5 & Code versioning system used & git \\
\hline
C6 & Software code languages, tools, and services used & Python, C, Make, GitHub Actions, cibuildwheel, PyPI \\
\hline
C7 & Compilation requirements, operating environments \& dependencies & Python $\geq$ 3.9, PyYAML. Pre-built binary wheels are provided for Linux x86$\_{64}$, macOS Intel, and macOS ARM64. \\
\hline
C8 & If available Link to developer documentation/manual & \url{https://rapiddocs.readthedocs.io/en/latest/} \\
\hline
C9 & Support email for questions & tarczaynehez.dora@csfk.org \\
\hline
\end{tabular}
\caption{Metadata of \texttt{rapidsim}.}
\label{} 
\end{table}

\section{Motivation and significance}

Scientific software plays a key role in investigating astrophysical phenomena, including star and planet formation and stellar evolution. 
Numerical simulations in astrophysics are often implemented in languages such as C, C++, and Fortran to meet their computational demands. 
However, installing and running these codes can require additional technical expertise and dependencies.
Users typically need to obtain the source code and its dependencies, configure the build environment, and compile the software before it can be used. 
This process may require platform-specific libraries, and careful configuration of include and library paths. 
Furthermore, differences between operating systems and computing environments can further complicate installation.

These challenges also arise, e.g., in the context of protoplanetary disc simulations, for example in the case of the \texttt{FARGO} and \texttt{FARGO3D} hydrodynamic codes \citep{Masset2000, fargo3d}, or when investigating stellar evolution by, e.g., \texttt{MESA} \citep[Modules for Experiments in Stellar Astrophysics][]{Paxton2011,Paxton2013,Paxton2015, Paxton2018, Paxton2019, Jermyn2023}.

\texttt{RAPID} \citep{TarczayNehez2026_rapid_first,Regalyetal2017} is a modular C simulation code developed to model the evolution of gas and dust in protoplanetary discs. 
The native executable of \texttt{RAPID} is constructed to run to computationally intensive one-dimensional hydrodynamic simulations. Although using the source code requires users to download the code from the GitHub repository\footnote{\url{https://github.com/tnehezd/RAPID}},  compile it and configure its dependencies in their local environment.
In order to simplify this process, \texttt{rapidsim} was developed as a Python package providing a higher-level interface to run \texttt{RAPID}.
It includes the \texttt{RAPID} executable in the package, allowing users to install and run simulations through the standard Python package manager without manually compiling the C source code.

Another important aspect of \texttt{rapidsim} is the management of software versions and distributions. 
Each release is linked to a specific version of \texttt{RAPID}, and binary distributions are built automatically for supported operating systems and Python versions. 
This approach simplifies installation and ensures that the distributed package contains the intended version of the simulation code.

\section{Software description}

\subsection{Software architecture}
\label{sec:arch}

The \texttt{rapidsim} package consists of a Python wrapper and the compiled \texttt{RAPID} simulation executable. The numerical calculations are performed by the native simulation engine, while the Python wrapper provides the interface between the user and the executable. The wrapper is implemented in Python and uses standard libraries for command-line argument parsing, file and path handling, subprocess management, and version reporting, together with PyYAML \citep{pyyaml} for reading the simulation configuration.

The wrapper organizes the simulation process into several steps. 
First, it reads the YAML \cite{pyyaml} configuration file and extracts the parameters specified in its sections. 
These sections cover the simulation setup, disk properties, boundary conditions, dead-zone parameters, dust properties, output settings, time integration, and logging. 
A parameter-mapping layer translates the configuration keys into the corresponding parameter names and command-line flags accepted by the \texttt{RAPID} executable. 
Boolean values are converted into numerical flags where required by the native interface, while logging settings determine the verbosity of the simulation output and whether terminal panels are disabled. 
This allows the simulation to be configured through a structured YAML file rather than by manually constructing a long command-line invocation.

Before launching the simulation, the wrapper searches for the executable in a predefined sequence of locations, including the binary bundled with the installed Python package and locations used in local development. It then constructs the command from the mapped parameters and launches the executable as a subprocess. 

The \texttt{RAPID} simulation code supports parallel execution using OpenMP. 
To configure the number of threads used during a simulation, \texttt{rapidsim} reads the \texttt{omp\_num\_threads} parameter from the YAML configuration file and sets the \texttt{OMP\_NUM\_THREADS} environment variable accordingly before launching the executable.
If this parameter is not specified, the wrapper preserves the existing environment setting, defaulting to a single thread if no value is defined. 
This allows users to control the degree of parallelism through the configuration file without having to set environment variables manually.
\texttt{rapidsim} also handles user interruption and terminates the subprocess when the simulation is cancelled.

The Python wrapper and the native simulation engine are maintained in separate GitHub repositories. The \texttt{RAPID} repository contains the numerical implementation and its native \texttt{make}-based build system, whereas the \texttt{rapidsim} repository contains the wrapper, packaging metadata, and release infrastructure. The repositories are linked through a Git submodule, with the \texttt{RAPID} source tree included in \texttt{rapidsim}. This arrangement keeps the development of the numerical code separate from that of the Python interface while allowing a specific version of the simulation engine to be associated with each package release.
When a new \texttt{RAPID} release is published, the automated release workflow builds a corresponding \texttt{rapidsim} package containing the source code of that \texttt{RAPID} release. The native executable is compiled using the \texttt{RAPID} build system and copied into the Python package's data directory. It is then distributed in the platform-specific binary wheel together with the Python wrapper. Consequently, users can install and run the package without obtaining or compiling the \texttt{RAPID} source code themselves, provided that a compatible pre-built wheel is available for their platform and Python version.

\subsection{Software functionalities}

The \texttt{rapidsim} wrapper simplifies the configuration and execution of \texttt{RAPID} by allowing users to specify simulation parameters in a YAML configuration file instead of providing them individually as command-line arguments. The configuration file uses a nested structure, grouping related parameters into sections covering the simulation setup, disk properties, boundary conditions, dead zone, dust, output, time integration, and logging. Figure~\ref{fig:config} shows an example configuration file illustrating this structure. A complete description of the available parameters and their accepted values is provided in the official online documentation.\footnote{\url{https://rapiddocs.readthedocs.io}}

The wrapper provides the following functionalities:
\begin{itemize}
    \item \textbf{Argument translation.} Configuration parameters are translated into the corresponding command-line arguments expected by the native executable. Boolean options are converted to the numerical flags required by the simulation engine, while logging settings are mapped to the appropriate verbosity switches.
    \item \textbf{Executable resolution.} The wrapper first searches for the executable bundled with the installed package. If it is not found, it falls back to checking \texttt{bin/simulation} relative to the package and the current working directory, allowing a locally built version of \texttt{RAPID} to be used during development.
    \item \textbf{Process control.} The simulation is launched as a subprocess, with its output streamed to the terminal in real time. The number of OpenMP threads is set according to the configuration, and the wrapper reports the process exit status. Pressing Ctrl+C interrupts the simulation.
    \item \textbf{Version reporting.} The \texttt{--version} option displays the installed \texttt{rapidsim} package version.
\end{itemize}

Precompiled native executables are distributed through platform-specific Python wheels. The release infrastructure builds wheels for Linux (x86$_{64}$) and macOS, supporting both Intel and Apple Silicon processors and multiple CPython versions. This enables package installers to select a compatible distribution for the user's platform and Python environment.

\subsection{Automated build and release system}

Binary distribution is automated using GitHub Actions and \texttt{cibuildwheel}. Publishing a new \texttt{RAPID} release triggers the \texttt{rapidsim} release workflow, which proceeds through the following steps:

\begin{enumerate}
    \item It identifies the new and previous \texttt{RAPID} release tags.
    \item It computes the next \texttt{rapidsim} version. The versioning of \texttt{rapidsim} is independent of that of \texttt{RAPID}, but follows its changes: an increase in the \texttt{RAPID} major version increments the \texttt{rapidsim} major version and resets the minor and patch numbers; an increase in the minor version increments the \texttt{rapidsim} minor version and resets the patch number; any other release increments the patch number.
    \item It checks out the corresponding \texttt{RAPID} release, compiles the native executable, and builds the platform-specific wheels.
    \item It verifies the contents of the generated packages and stores the distributions as release artifacts.
    \item After all target platforms have been built successfully, it commits the updated submodule reference and version metadata to the repository and publishes the wheels to PyPI.
\end{enumerate}

\begin{figure}
    \centering
    \includegraphics[width=\linewidth]{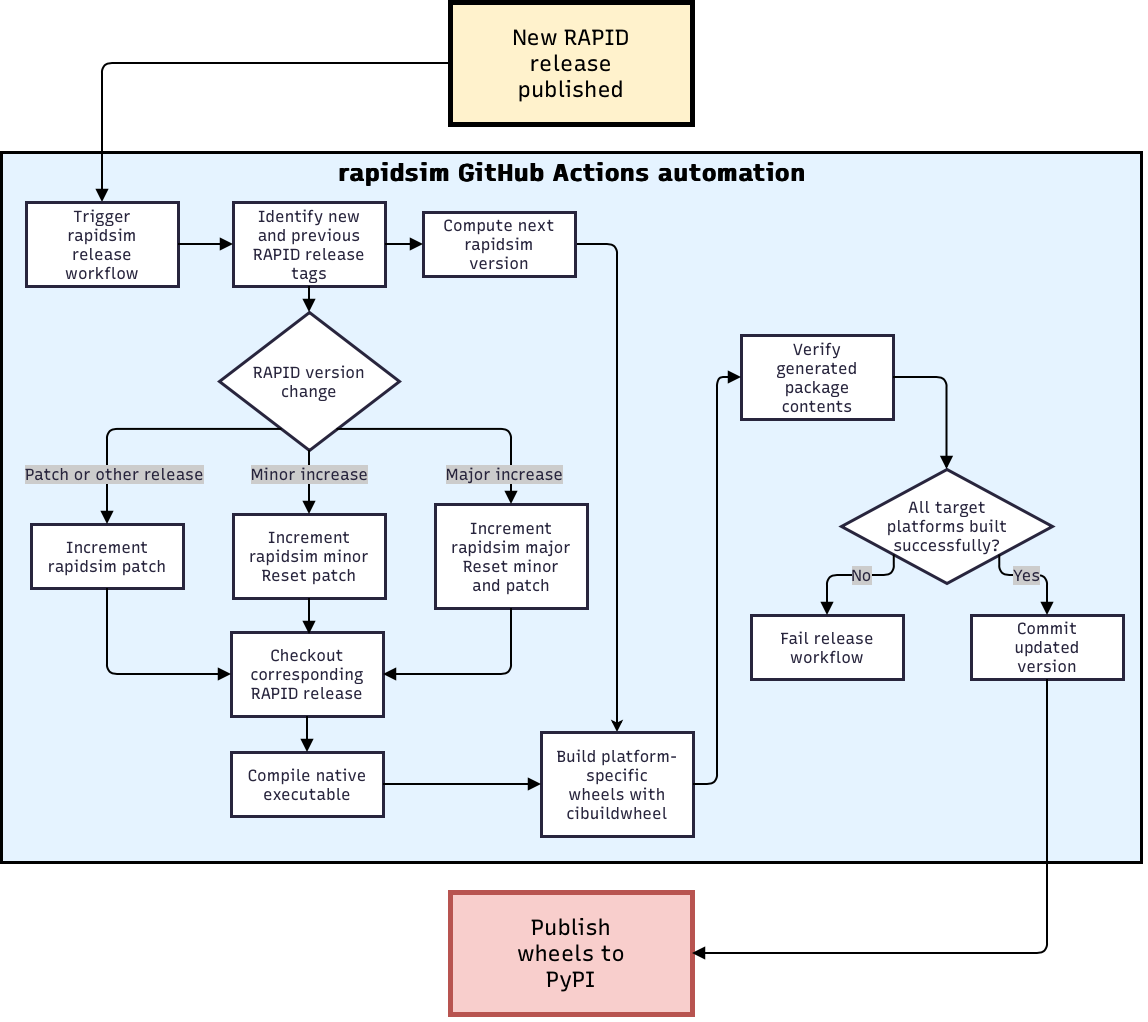}
    \caption{Flowchart of the GitHub actions automation of publishing \texttt{rapidsim} upon new \texttt{RAPID} version release.}
    \label{fig:automaton}
\end{figure}

This automation removes manual steps from the release process. 
Each published package is built from a specific revision of the \texttt{RAPID} source tree, and the resulting binary distributions are generated and published through a common workflow.
Figure~\ref{fig:automaton} presents a flowchart of the automated publish process of \texttt{rapidsim}.

\section{Illustrative examples}

\subsection{Installing \texttt{rapidsim}}

The \texttt{rapidsim} package requires at least a Python 3.9 environment. It can be installed using the standard Python package manager:

\begin{verbatim}
   > pip install rapidsim
\end{verbatim}

The project is hosted on GitHub as an open-source code, for development mode, the package can also be cloned as:

\begin{verbatim}
   > git clone https://github.com/tnehezd/rapidsim.git
\end{verbatim}

The package can then be installed in editable mode, allowing modifications to the Python wrapper. 
Such modifications can be used to customize the configuration handling or execution workflow. 
Adding new simulation parameters or command-line switches, however, requires corresponding changes to the native \texttt{RAPID} code.

The package also supports installation on heterogeneous systems: \texttt{pip} automatically selects the appropriate binary wheel for the host platform and Python version. 
This was tested on Intel-based macOS and Linux systems, including a Linux environment running Python 3.13, demonstrating that the package can be installed and executed outside its original development environment.

\subsection{Running \texttt{rapidsim}}

After installation, a simulation can be launched using the \texttt{rapidsim} command-line interface together with a \texttt{RAPID} configuration file:

\begin{verbatim}
   > rapidsim --config path/to/config.yaml
\end{verbatim}

The configuration file should follow a structured format.
Figure~\ref{fig:config} shows an example of a typical configuration file for a simple, pure gas disk.
A more detailed description of the available configuration parameters and their accepted values can be found in the package documentation and in the defaults directory of the package's GitHub repository.

Figure~\ref{fig:rapid_init_output}  present a typical initialization output on the terminal for a pure gas evolving disk.
When enabled, an update panel shows the current status of the simulation (see Figure~\ref{fig:main_loop}).
Upon successful completion, \texttt{rapidsim} reports the end of the simulation.
The output files are saved in the user-specified format (HDF5 or ASCII) in the designated output directory. 
If the default \texttt{output} directory already exists, an incremented four-digit suffix (starting from \texttt{0000}) is appended to avoid overwriting previous results. 
The output directory contains the initial-condition files for the disk and dust, while the snapshots are stored in the \texttt{LOGS} directory or a user-defined alternative. 
Additional log files, including performance logs, are also saved in the output directory.

\begin{figure}
\begin{lstlisting}[
  frame=single,
  framesep=8pt,
  rulecolor=\color{black!30},
  backgroundcolor=\color{gray!5},
  basicstyle=\ttfamily\scriptsize,
  showstringspaces=false,
  breaklines=true
]


[PY-WRAPPER: RAPIDSIM] --- Running: Main Simulation Program ---
[PY-WRAPPER: RAPIDSIM] The current command-line arguments are:
 simulation -drift 0.0 -growth 0.0 -evol 1.0 -photoevap 0.0 -grid_type linear -n 500 -ri 1 -ro 50 -disk_mass 0.01 -index_init -1.0 -alpha_init 0.01 -inner_bc zero_gradient -outer_bc zero_gradient -rdzei 0 -rdzeo 0 -amod 0 -o output_gas_dze -output_format ascii -tmax 1e5 -outfreq 10000
[PY-WRAPPER: RAPIDSIM] Setting OMP_NUM_THREADS=1 for this run.
[PY-WRAPPER: RAPIDSIM] Start running the binary

...

==========================================================================
||                                                                      ||
||                        RAPID SIMULATION CORE                         ||
||                            Version: 2.2.7                            ||
||                                                                      ||
||                  Lead Developer: [D. Tarczay-Nehez]                  ||
||                      Co-Developer: [A. Csaszar]                      ||
||                                                                      ||
==========================================================================

               Initializing simulation core...

==========================================================================
||                      INITIALIZATION PARAMETERS                       ||
==========================================================================
    Stellar mass (Solar Mass):          1
    Disk aspect ratio (H/R):            0.05
    Disk flaring index:                 0
    Alpha viscosity:                    0.01
    Total disk mass (Solar Mass):       0.01
    Inner disk edge (AU):               1
    Outer disk edge (AU):               50
    Profile Style:                      [Standard Pure Power-Law]
    Surface density profile exponent:   1
    Gas surface density at 1 AU:        3.24806e-05 M_Sun/AU^2
    Number of gas grid points:          500
==========================================================================        

...

==========================================================================
||                         BOUNDARY CONDITIONS                          ||
==========================================================================
    INNER BC:       [zero]
    OUTER BC:       [parabolic]
==========================================================================

...
                      
INITIALIZATION COMPLETE.

        
\end{lstlisting}
\caption{Typical initialization terminal output of the \texttt{RAPID} simulation core for a pure gas accreting model.The \texttt{[PY-WRAPPER: RAPIDSIM]} label indicates that the simulation is launched via \texttt{rapidsim}.}
\label{fig:rapid_init_output}
\end{figure}

\begin{figure*}
\begin{lstlisting}[
  frame=single,
  framesep=8pt,
  rulecolor=\color{black!30},
  backgroundcolor=\color{gray!5},
  basicstyle=\ttfamily\scriptsize,
  showstringspaces=false,
]
   

==========================================================================
||                   STARTING MAIN SIMULATION LOOP...                   ||
==========================================================================

[SAVE AT STEP:      1] Time: 0.00e+00 yr | dt: 1.91e+00 yr | HDF5 SNAPSHOT SAVED
==========================================================================
| STEP:        30010    | Mode: GAS-ONLY EVOLUTION
| TIME:        9.9993e+03 / 1.0000e+04 yr  [###################.] 100.0%
| DT:          1.9091e+00 yr
| DISK MASS:   1.0020e-02 / 1.0020e-05 M_Sun (Remaining: 100.00%)
==========================================================================

        ...


[SAVE AT STEP: 300200] Time: 9.00e+04 yr | dt: 1.91e+00 yr | HDF5 SNAPSHOT SAVED
==========================================================================
| STEP:        300200   | Mode: GAS-ONLY EVOLUTION
| TIME:        9.9999e+04 / 1.0000e+05 yr  [###################.] 100.0%
| DT:          1.9091e+00 yr
| DISK MASS:   1.0020e-02 / 1.0020e-05 M_Sun (Remaining: 100.00%)
==========================================================================


==========================================================================
||      Simulation finished in 00:01:32. Summary written to logs.       ||
==========================================================================

[PY-WRAPPER: RAPIDSIM] Main Simulation Program completed successfully.

\end{lstlisting}
\caption{Example progress and finishing panels for a pure gas simulation initialized in Figure~\ref{fig:rapid_init_output}.}
\label{fig:main_loop}
\end{figure*}

\section{Impact}

The primary impact of \texttt{rapidsim} is the reduction of the technical barrier between a computational astrophysics code and its end users.
Without the package, users of \texttt{RAPID} must obtain the source code, satisfy the compiler and native library requirements, and build the executable locally.
\texttt{rapidsim} transfers this complexity to an automated build infrastructure and presents the resulting software through a standard Python distribution mechanism.
By retaining the computational core in C, \texttt{rapidsim} combines the performance advantages of compiled native code with the ease of installation and execution provided by a Python interface. 
This approach is widely used in the scientific Python ecosystem, where packages such as NumPy \citep{numpy} expose computationally intensive native implementations through a high-level Python interface.

Furthermore, the separation between \texttt{RAPID} and \texttt{rapidsim} establishes a reusable distribution model for future development. 
\texttt{RAPID} can be developed as a native numerical code while \texttt{rapidsim} maintains a stable user-facing interface and provides reproducible binary distributions. 
Conversely, modifications to the Python interface and packaging infrastructure do not require changes to the numerical implementation,  unless new command-line options or simulation parameters need to be introduced.
The automated release system further improves the traceability of the distributed software. 
Each package release is constructed from a specific \texttt{RAPID} release, with the source revision recorded automatically and the corresponding binary packages published to PyPI.

Although \texttt{rapidsim} has not yet been the subject of a dedicated scientific publication, the underlying \texttt{RAPID} code has already been used in published research papers \citep{Regalyetal2017, TarczayNehez2026_rapid_first}.
 Additionally, Tarczay-Nehéz~\&~Császár \cite{tarczaynehezcsaszar2026rapidmethods} presents a dedicated methods paper describing the code design, numerical methods, and implementation of  \texttt{RAPID} in detail.
 Furthermore, an extension of \cite{TarczayNehez2026_rapid_first} is in preparation to explore additional aspects of protoplanetary disk evolution, including photoevaporation. The availability of \texttt{rapidsim} facilitates these follow-up investigations by simplifying software installation and simulation execution while retaining the existing computational capabilities of \texttt{RAPID}.

\section{Conclusions}

\texttt{rapidsim} provides a Python interface and cross-platform distribution for \texttt{RAPID}, an established C-based code for simulating gas and dust evolution in protoplanetary discs. 
The main contribution of \texttt{rapidsim} is to make the existing numerical capabilities of \texttt{RAPID} accessible through a standard Python package, without requiring users to compile the native code or manage its build dependencies themselves.
The functionality provided by \texttt{rapidsim} includes YAML-based control of simulation parameters, pre-built binary wheels for supported Linux and macOS platforms, and an automated release pipeline that links each distribution to a specific version of the underlying simulation code.
This separation of the native C code \texttt{RAPID} and the wrapper \texttt{rapidsim} allows the numerical implementation and the user-facing interface to evolve independently, while maintaining traceability between source code and distributed executables.
The resulting package provides a practical foundation for using \texttt{RAPID} in a wider range of research environments and for incorporating its simulations into Python-based scientific workflows.

\section*{Acknowledgements}

This paper was supported by the János Bolyai Scholarship of the Hungarian Academy of Sciences, by the `SeismoLab' KKP-137523 \'Elvonal grant of the Hungarian Research, Development and Innovation Office (NKFIH), and by the LP2025-14/2025
Lendület grant of the Hungarian Academy of Sciences.

This paper is written by the assistance of a large language model (ChatGPT-5.6 Luna by OpenAI) for linguistic  refinement. 
All content and any errors remain the sole responsibility of the first author.

\appendix

\section{An example of a YAML configuration file}
\label{app:example_yaml}

\lstdefinelanguage{yaml}{
  keywords={
    simulation_parameters, disk_parameters, boundary_conditions, 
    deadzone_parameters, dust_parameters, output_parameters, 
    time_parameters, log_parameters
  },
  keywordstyle=\color{green!60 !black}\bfseries,
  emph={true, false, null},
  emphstyle=\color{blue}\bfseries,
  alsoother={:},
  basicstyle=\ttfamily,
  string=[b]",
  stringstyle=\color{orange!80!black}\bfseries,
  sensitive=false,
  comment=[l]{\#}
}

\begin{figure}
    \centering
\begin{lstlisting}[
  language=yaml,
  frame=single,
  framesep=8pt,
  rulecolor=\color{black!30},
  backgroundcolor=\color{gray!5},
  basicstyle=\ttfamily\footnotesize,
  showstringspaces=false,
  breaklines=true
]
# example YAML config file

simulation_parameters:
  # Simulation Control Options 
  enable_dust_drift:           false     
  enable_dust_growth:          false     
  enable_gas_evolution:        true    
  enable_two_dust_populations: false 
  enable_photoevaporation:     false      
  
disk_parameters:
  # Grid and Disk Initial Parameters 
  number_of_grid_points:   500    
  inner_radius_au:         1           
  outer_radius_au:         50.0         
  disk_mass:               0.01              
  sigma_profile_exponent: -1.0  
  alpha_viscosity:         0.01      
  aspect_ratio_at_1au:     0.05    

output_parameters:
  # File I/O Parameters
  output_directory_name: "output"    
  output_format:         "ascii"               

time_parameters:
  # Time Parameters
  total_simulation_time:  1e5       
  output_write_frequency: 10000
  
\end{lstlisting}
    \caption{An example of a YAML configuration file for running simulation with \texttt{rapidsim}.}
    \label{fig:config}
\end{figure}

\bibliographystyle{elsarticle-num} 
\bibliography{sample}






\section*{Current executable software version}
\label{}
The current version of \texttt{rapidsim} is v1.0.15.

\end{document}